\documentclass{jfm}
\usepackage{graphicx}
\usepackage{newtxtext}
\usepackage{newtxmath}
\usepackage{natbib}
\usepackage{hyperref}
\usepackage{cleveref}
\crefformat{section}{\S#2#1#3} 
\crefformat{subsection}{\S#2#1#3}
\crefformat{subsubsection}{\S#2#1#3}
\hypersetup{
    colorlinks = true,
    urlcolor   = blue,
    citecolor  = black,
}

\newcommand{\RomanNumeralCaps}[1]
\linenumbers

\shorttitle{Evidence of preferential sweeping during snow settling in atmospheric turbulence}
\shortauthor{J. Li, A. Abraham, M. Guala and J. Hong}

\title{Evidence of preferential sweeping during snow settling in atmospheric turbulence}

\author{Jiaqi Li\aff{1,2},
  Aliza Abraham\aff{1,2},
  Michele Guala\aff{1,3}
 \and Jiarong Hong\aff{1,2}
   \corresp{\email{jhong@umn.edu}}}

\affiliation{\aff{1}Saint Anthony Falls Laboratory, University of Minnesota, Minneapolis, MN, 55455, USA
\aff{2}Department of Mechanical Engineering, University of Minnesota, Minneapolis, MN, 55455, USA
\aff{3}Department of Civil, Environmental and Geo- Engineering, University of Minnesota, Minneapolis, MN, 55455, USA}

\begin{document}

\maketitle

\begin{abstract}
We present a field study of snow settling dynamics based on simultaneous measurements of the atmospheric flow field and snow particle trajectories. Specifically, a super-large-scale particle image velocimetry (SLPIV) system using natural snow particles as tracers is deployed to quantify the velocity field and identify vortex structures in a 22 m $\times$ 39 m field of view centered 18 m above the ground. Simultaneously, we track individual snow particles in a 3 m $\times$ 5 m sample area within the SLPIV using particle tracking velocimetry (PTV). The results reveal the direct linkage among vortex structures in atmospheric turbulence, the spatial distribution of snow particle concentration, and their settling dynamics. In particular, with snow turbulence interaction at near-critical Stokes number, the settling velocity enhancement of snow particles is multifold, and larger than what has been observed in previous field studies. SLPIV measurements show higher concentration of snow particles preferentially located on the downward side of the vortices identified in the atmospheric flow field. PTV, performed on high resolution images around the reconstructed vortices, confirms the latter trend and provides statistical evidence of the acceleration of snow particles, as they move toward the downward side of vortices. Overall, the simultaneous multi-scale particle imaging presented here enables us to directly quantify the salient features of preferential sweeping, supporting it as an underlying mechanism of snow settling enhancement in the atmospheric surface layer. 
\end{abstract}


\section{Introduction}
\label{sec:intro}
Understanding the settling dynamics of inertial particles in turbulence is important for predicting particle transport in the atmosphere, such as aeolian transport of dust and sand \citep{Duran11}, formation and growth of droplets and particle aggregates in clouds \citep{Shaw03}, and precipitation of hydrometers, such as raindrops, graupels and snowflakes \citep{Garrett15,Nemes17,Zeugin20,Li21}. Numerous laboratory experiments and numerical simulations have been conducted to investigate the effects of turbulence on the behavior of inertial particles. Two evident manifestations of particle-turbulence interaction mechanisms are the formation of particle clusters and the modulation of their settling velocity \citep{Balachandar10}. These phenomena are observed in certain conditions depending on the turbulence (through the Kolmogorov time scale, $\tau_{\eta}$), and on the particle size (${\it D_p}$), density ($\rho_p$) and aerodynamic properties, contributing to the definition of the particle response time, $\tau_p$ \citep{Maxey83}. The phenomena of clustering and enhanced settling can be described as follows: as particles preferentially concentrate in strain-dominated regions (e.g., in between vortices), they settle along the downward side of swirling motions as clusters. As a result, the fall speed of the particles on the downward side is increased. This mechanism is known as preferential sweeping \citep{Wang93}. Studies have shown that the average settling velocity of inertial particles in turbulence can be enhanced significantly by the preferential sweeping mechanism \citep{Wang93,Yang98, Aliseda02,Good14,Falkinhoff20}, in particular under critical conditions, i.e. when the Stokes number $\Stk = \tau_p/\tau_{\eta} \approx 1$ \citep{Yang98,Aliseda02,Ferrante03}. There are also other mechanisms that have been described to hinder the settling of inertial particles in turbulence such as loitering \citep{Nielson93} and vortex trapping \citep{Tooby77}, but they usually tend to be suppressed by preferential sweeping \citep{Good14,Rosa16}. 

Despite the large number of laboratory experiments and simulations, field measurements of inertial particles (e.g. snow particles, droplets, and dust) settling in the atmospheric turbulence are scarce. The lack of field evidence is mostly due to the fact that field measurements are experimentally challenging \citep{Shaw03}: local turbulent field conditions are difficult to parameterize, and the effects of particle interaction and flow Reynolds numbers on non-Stokesian particle kinematics is far from being clear (see recent advancements by \citet{Petersen19}, \citet{Tom19}, and \citet{Falkinhoff20}). Moreover, the implementation of particle-turbulence interaction mechanisms in predictive models of settling velocity at geophysical scales is also limited, since the field conditions (e.g. wide range of turbulence scales, complex particle shape) are often different from those reproduced in laboratory experiments and simulations.

To enable spatially and temporally resolved flow measurements in the field, a super-large-scale particle image velocimetry (SLPIV), using natural snow particles as tracers, has been recently developed for studying the wake structure downstream of a utility scale wind turbine in the atmospheric boundary layer \citep{Hong14,Dasari19,Abraham20} and for the study of high Reynolds number wall turbulence \citep{Toloui14,Heisel18}. Using the similar setup, \citet{Nemes17} quantified the settling trajectories of ~87000 snow particles in a 4 m (width) $\times$ 7 m (height) field of view using particle tracking velocimetry (PTV), in parallel with a digital inline holography (DIH) system, to characterize the size and morphology of snow particles. In the absence of direct estimates of snow particle density, the acceleration probability density function (PDF) obtained by PTV was used to estimate the Stokes number and the aerodynamic particle response time of snow particles \citep{Mordant04,Bec06,Ayya06}. \citet{Nemes17} found that the settling velocity of snow particles measured using PTV showed multifold enhancements in the atmospheric turbulence, in comparison to the still-air terminal velocity \textbf{${\it W_p}= \tau_p\cdot{g}$} predicted using the acceleration-based aerodynamic response time. Employing the same setup, \citet{Li21} investigated the settling and clustering of snow particles under various turbulence and snow conditions. They observed intense clustering and enhanced settling velocity during near-critical Stokes conditions, showing statistical evidence of the correlation between enhanced settling velocity and local preferential concentration, thus indirectly supporting the preferential sweeping mechanism. Despite these major findings, both contributions could not simultaneously provide flow measurements and trajectories of snow particles.

In the present study, we leverage the ability of SLPIV to measure large scale flows and PTV to observe and track individual particles, allowing us to unveil the direct, local, linkages between coherent vortex structures, snow concentration distribution, and settling velocity at $\Rey_\lambda\sim\textit{O}(10^3)$. We quantify here both the preferential concentration around vortices and enhanced settling velocity on the downward side of vortices, thus highlighting the fundamental mechanism of preferential sweeping. The experiment setup, atmospheric conditions, and turbulence properties are introduced in \cref{sec:method}. In \cref{sec:results}, results are presented for the quantification and visualization of preferential sweeping mechanism. Conclusions and discussion follow in \cref{sec:con}.

\begin{figure}
  \centerline{\includegraphics[scale = 0.85]{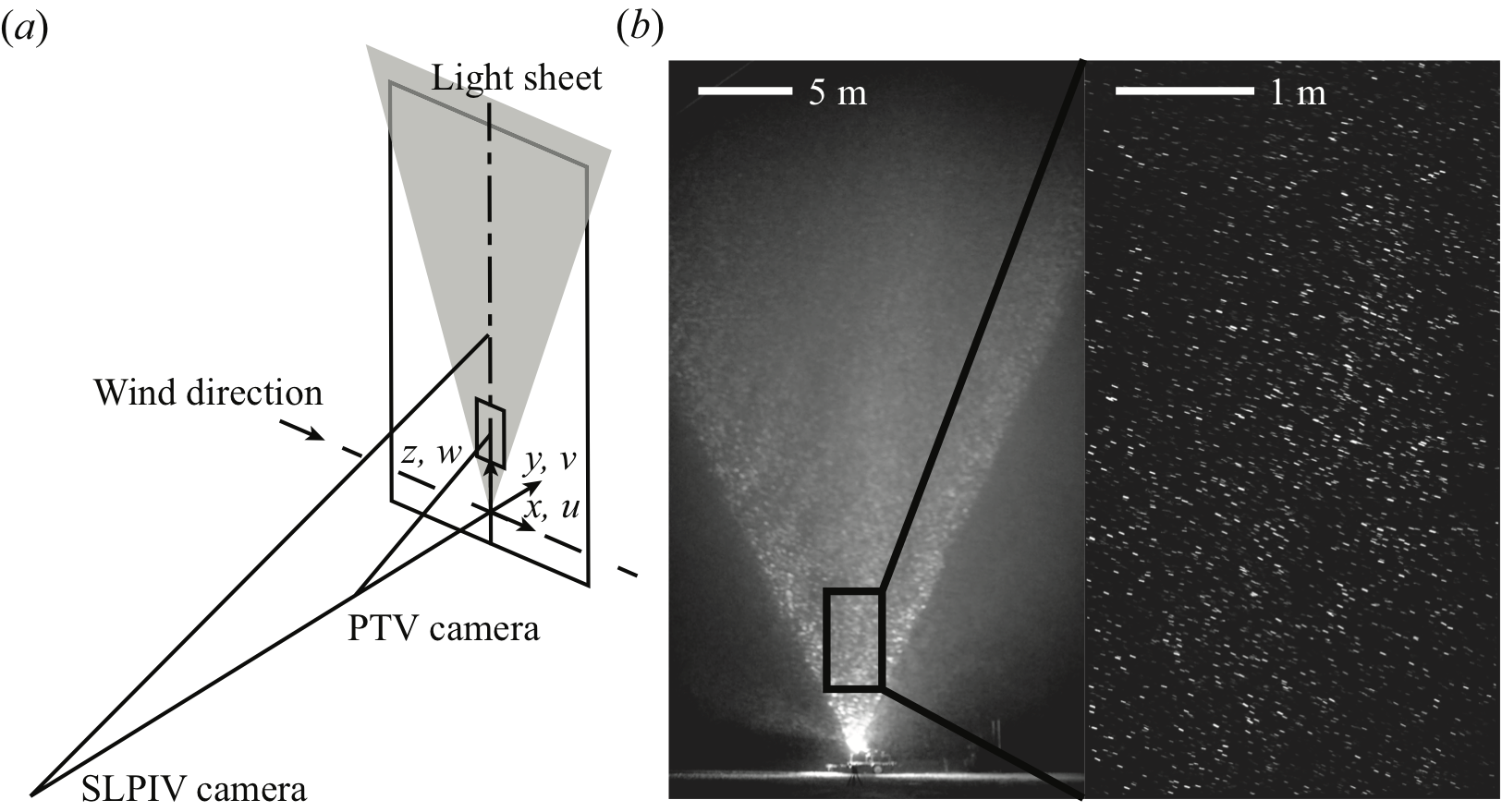}}
  \caption{(\textit{a}) The experiment setup of simultaneous super-large-scale PIV (SLPIV) and PTV measurements; (\textit{b}) sample images showing the fields of view of SLPIV (left) and PTV (right).}
\label{fig:1}
\end{figure}

\section{Methodology}\label{sec:method}
\subsection {Field experiment setup}\label{sec:21}
The field deployment was conducted to acquire data on Jan. 13, 2020 between 18:00 and 21:00 local time, at the Eolos Wind Energy Research Field Station in Rosemount, MN. The light sheet-based super-large-scale particle image velocimetry (SLPIV) and particle tracking velocimetry (PTV), described in \citet{Hong14} and \citet{Nemes17}, respectively, have been applied to capture the turbulent flow field, the trajectories and the concentration distribution of snow particles. We used a 5-kW search light with a curved mirror expanding the beam vertically into a light sheet to illuminate the snow particles. For our current measurements, the light sheet thickness is restricted to be 10 cm (different from our previous measurements with 30 cm diameter light beam) at the ground and it increases to about 12 cm at 10 m considering the divergence angle of our search light. The light sheet was oriented to be parallel with the average wind direction and minimize the out-of-plane motion. Throughout the deployment, the instantaneous wind direction relative to the light sheet varied from -25 degrees to 15 degrees. An 11-minute duration dataset has been selected for the measurement presented in this paper: within the selected period of time the wind direction was stable and well-aligned with the light sheet direction with a deviation of less than five degrees, and the snowfall intensity was steady, providing adequate seeding density of around 150 snow particles per $\textnormal{m}^2$.

A Nikon D600 (Nikon Inc.) camera and a Sony A7RII (Sony Corp.) camera were equipped to record the overall flow field at 30 fps and 1080 $\times$ 1920 $\textnormal{pixel}^2$, and the motion of snow particles at 120 fps and 720 $\times$ 1280 $\textnormal{pixel}^2$ respectively (referred as SLPIV and PTV in the following sections). The start time of each recording for both cameras is documented, and a large-scale turbulent structure visible as a void at the beginning of both videos is used to further confirm the synchronization of the two datasets. Both cameras were placed on tripods with measured tilt angles from the horizontal direction (table \ref{tab:1}). The relative locations of the two cameras and the light sheet are illustrated in figure \ref{fig:1}a with the defined coordinate system (\textit{x}, \textit{y}, \textit{z}) and corresponding velocity components (\textit{u}, \textit{v}, \textit{w}). The specifications (e.g. duration, field of view (FOV) elevation, size and distance to the camera, etc.) for the two cameras are shown in table \ref{tab:1}.

\begin{figure}
  \centerline{\includegraphics[scale = 0.85]{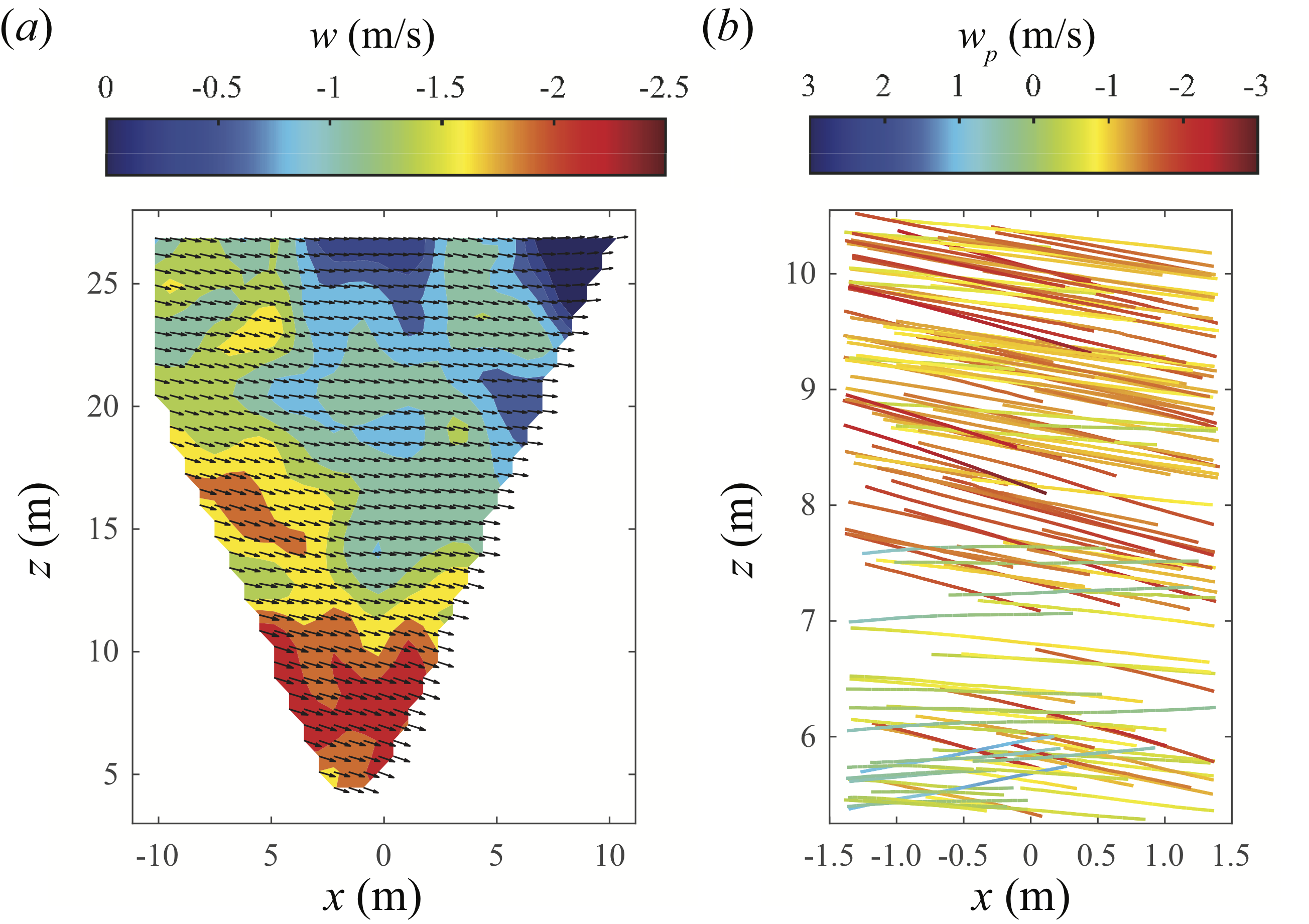}}
  \caption{(\textit{a}) Instantaneous flow field sample from SLPIV; (\textit{b}) and snow particle trajectory samples from PTV.}
\label{fig:2}
\end{figure}

Sample images for SLPIV (Nikon camera) and PTV (Sony camera) are shown in figure \ref{fig:1}b, and sample results are shown in figure \ref{fig:2}. PIV analysis is conducted using LaVision Davis 8.2.0. A multi-pass setting was adopted with a final pass of 32 $\times$ 32 $\textnormal{pixel}^2$ and 50$\%$ overlap. Around 700 vectors are obtained from each image pair. For PTV, we apply the learning-based tracking method from \citet{Mallery20} using a long short-term memory (LSTM) network to acquire individual trajectories. Specifically, we first implement tracking methods from \citet{Crocker96} and \citet{Ouellette06} to our PTV data. However, due to the lower quality of field data and relatively high particle concentration, the conventional methods generate substantially less tracks compared to those can be determined through manual examination, potentially causing sampling bias (i.e., preferentially shorter tracks, sampling only downward tracks). Therefore, good quality trajectories generated by the conventional methods are manually selected as the training set for the learning-based method. After iterations of training process, the well-trained model generates significantly more tracks regardless of their direction (upward or downward). In total, there are around 460,000 trajectories longer than $10\tau_\eta$ (where $\tau_\eta$ the Kolmogorov time scale) being identified by the tracking algorithm. After the flow field and trajectories are obtained, we further identify the distribution of potential vortical structures in the flow field based on swirling strength \citep{Zhou99} and calculate the Lagrangian velocity and acceleration using the trajectory information. The swirling strength is defined as $\lambda_{ci}$, the imaginary part of the complex eigenvalues of the velocity gradient tensor ($\textbf{\textit{D}}=\nabla\textbf{\textit{u}}$). Under two-dimensional measurement, $\textbf{\textit{D}}$ have either two real eigenvalues ($\lambda_{r}$) or a pair of conjugate complex eigenvalues ($\lambda_{cr} \pm i \lambda_{ci}$), where $\lambda_{cr}$ and $\lambda_{ci}$ are absolute values. Thus, the vortices can be identified with finite $\lambda_{ci}$ \citep{Adrian00}. Three threshold values of the swirling strength (0.4 $\textnormal{s}^{-1}$, 0.5 $\textnormal{s}^{-1}$ and 0.65 $\textnormal{s}^{-1}$) are applied for detecting the vortices, and the concentration and settling velocity are analyzed using all three threshold values. In the results section, we will show the preferential concentration and enhanced settling with the 0.4 $\textnormal{s}^{-1}$ threshold, and figures with the other thresholds will be shown in the supplementary material. 

Following the same procedures as in our previous studies \citep{Toloui14,Hong14}, the traceability of snow particles for our SLPIV measurement is analyzed. Specifically, the spatial resolution of the SLPIV is usually limited by the smallest interrogation window size and light sheet thickness (whichever is larger), i.e., $l=0.66$ m in our current SLPIV measurements. Correspondingly, the flow time scale that our measurements can resolve is estimated as $\tau_{f} = l/u_{\textnormal{rms}}=0.62$ s, where $u_{\textnormal{rms}} = 1.07$ m/s is the r.m.s. of the streamwise velocity fluctuations. Thus, the particle Stokes number based on the particle response time ($\tau_{p} = 1.7-20$ ms, from acceleration PDF in \cref{sec:31}) and $\tau_{f}$ is estimated to be $\Stk = \tau_{p} / \tau_{f} = 0.0028-0.032$, much smaller than the typical threshold for good traceability (i.e., 0.1 according to \citet{Tropea07}). As a result, turbulent flows above the spatial and temporal resolution limits (i.e., 0.66 m and 0.62 s, respectively) are reasonably captured in our measurements. Note also that the mean settling velocity of the snow particles is subtracted from the SLPIV flow field.

A digital inline holography (DIH) \citep{Nemes17} system was deployed, and around 1580 snow particles are captured during the 11 min of the SLPIV and PTV data. The mean snow particle equivalent diameter is measured to be 0.39 mm with a standard deviation of 0.29 mm, and the average aspect ratio of the fitted minor and major ellipsoid axis is 0.62. The sample volume for the DIH measurement is 42 $\textnormal{cm}^3$, thus the mean snow particle number concentration is around 28,460 $\textnormal{m}^{-3}$, and the volume fraction is $3.8 \times 10^{-6}$.

\begin{table}
  \begin{center}
\def~{\hphantom{0}}
  \begin{tabular}{ccccccc}
  \multicolumn{7}{c}{SLPIV/PTV setup} \\
  \hline
      \multicolumn{1}{c}{\begin{tabular}[c]{@{}c@{}}Deployment \\ datasets \end{tabular}}  & \multicolumn{1}{c}{\begin{tabular}[c]{@{}c@{}}Duration \\ (min) \end{tabular}} & \multicolumn{1}{c}{\begin{tabular}[c]{@{}c@{}}Elevation \\ (m) \end{tabular}} & \multicolumn{1}{c}{\begin{tabular}[c]{@{}c@{}}FOV size \\ ($\textnormal{m}^2$) \end{tabular}} & \multicolumn{1}{c}{\begin{tabular}[c]{@{}c@{}}Resolution \\ (mm/pixel) \end{tabular}} & \multicolumn{1}{c}{\begin{tabular}[c]{@{}c@{}}Camera angle \\ (deg.) \end{tabular}} & \multicolumn{1}{c}{\begin{tabular}[c]{@{}c@{}}Camera-to-light \\ distance (m) \end{tabular}} \\[5pt]
       SLPIV & 11 & 18.4 & 39.2 $\times$ 22.1 & 20.5 & 19.9 & 52.5 \\
       PTV   & 11 & 7.9 & 5.3 $\times$ 3.0 & 4.15 & 18.4 & 19.1 \\
  \end{tabular}
  \caption{Summary of key parameters of the SLPIV and PTV measurement setups for the deployment dataset used in the present paper.}
  \label{tab:1}
  \end{center}
\end{table}

\begin{table}
  \begin{center}
\def~{\hphantom{0}}
  \begin{tabular}{cccccccccccccc}
      \textit{U} & $u_\textnormal{rms}$ & $w_\textnormal{rms}$ & $R_b$ & $L_\textnormal{OB}$ & \textbf{\textit{$U_{\tau}$}} & \textbf{\textit{$z/L_\textnormal{OB}$}} & \textit{L} & $\tau_L$ & $\epsilon$ & $\eta$ & $\tau_\eta$ & $\lambda$ & $\Rey_\lambda$ \\
      m/s & m/s & m/s & - & m & m/s & - & m & s & $\textnormal{cm}^2/\textnormal{s}^3$ & mm & ms & mm & - \\[3pt]
      5.47 & 1.07 & 0.64 & 0.16 & 1651 & 0.48 & 0.0062 & 6.22 & 5.79 & 355 & 0.51 & 19.4 & 80.7 & 6478 \\
  \end{tabular}
  \caption{Estimated meteorological and turbulence conditions from the sonic anemometer at $z = 10 \textnormal{m}$. See the text for the definition of the symbols.}
  \label{tab:2}
  \end{center}
\end{table}

\subsection{Atmospheric turbulence conditions}
The atmospheric and turbulence conditions during the deployment are determined using a meteorological tower instrumented with wind velocity, temperature and humidity sensors at 50 m downstream of the light sheet. Four sonic anemometers (20 Hz sampling rate, 5.8 cm horizontal measurement path length and 10 cm vertical measurement path length, CSAT3, Campbell Scientific) are installed at elevations of 10, 30, 80 and 129 m, and six cup-and-vane anemometers (1 Hz sampling rate) are installed at elevations of 7, 27, 52, 77, 102 and 126 m. Note that the measurement uncertainty of the sonic anemometer is $\pm$0.08 m/s \citep{Toloui14}, corresponding to 1.4$\%$ of the average wind speed. Thus, we estimate that the uncertainties for the turbulence properties would be less than 4$\%$. The key parameters are listed in table \ref{tab:2}. The atmospheric stability is estimated using the bulk Richardson number $R_b$ and the Monin-Obukhov length $L_\textnormal{OB}$:

\begin{equation}
R_{b}=-|g| \Delta \overline{\theta_{v}} \Delta z /\left(\overline{\theta_{v}}\left[\left(\Delta V_{N}\right)^{2}+\left(\Delta V_{W}\right)^{2}\right]\right)
\end{equation}

\begin{equation}
L_{O B}=-U_{\tau}^{3} \overline{\theta_{v}} / \kappa g \overline{w^{\prime} \theta_{v}^{\prime}}
\end{equation}

In the equations, \textit{g} is the gravitational acceleration; $\theta_{v}$ is the virtual potential temperature; $V_N$ and $V_W$ are the average wind velocity components to the North and West respectively measured by the sonic anemometers; $\kappa$ is the von Kármán constant; $U_{\tau}$ is the shear velocity estimated from the Reynolds stresses \citep{Stull88}, where $U_{\tau}=\left(\left\langle V_N^{\prime} V_Z^{\prime}\right\rangle^{2}+\left\langle V_W^{\prime} V_Z^{\prime}\right\rangle^{2}\right)^{1/4}$. The average velocity differences are calculated from the sonic anemometers at top (129 m) and bottom (10 m) which yields a height difference $\Delta\textit{z}$ of 119 m. The Monin-Obukhov length and all other turbulence conditions are measured with the data from the 20 Hz sonic sensor at 10 m. For the duration of the analyzed dataset, the bulk Richardson number and the Monin-Obukhov length indicate that the atmospheric boundary layer during the experiment is near-neutrally stratified (typically for the near neutrally stratified atmospheric boundary layer, $R_b$ ranges from 0 to 0.25 and stability parameter ($z/L_\textnormal{OB}$) ranges from 0 to 0.1 \citep{Hogstrom02, Stull88}).

The turbulence conditions are estimated using the methods described in \citet{Nemes17} and \citet{Li21}. Velocity data from the sonic anemometer at 10 m are used for the flow characterization, consistent with the sample area elevation ranges of the SLPIV (from ~3 m to around 40 m) and PTV (from 5.3 m to 10.6 m) measurements.The integral time scale $\tau_{L}$ and the length scale \textit{L} are estimated based on the temporal autocorrelation function $\rho_{u u}$:

\begin{equation}
\rho_{u u}(\tau)=\left\langle u^{\prime}(t) u^{\prime}(t+\tau)\right\rangle / u^{\prime 2}
\end{equation}

\begin{equation}
\tau_{L}=\int_{0}^{T_{0}} \rho_{u u}(\tau) d \tau
\end{equation}

\begin{equation}
L=u_\textnormal{rms} \tau_{L}
\end{equation}

In these equations, \textit{t} is the variable time, $\tau$ is the time difference, and $T_0$ is the first zero-crossing point the auto-correlation function. The turbulence dissipation rate $\epsilon$ is estimated using the second-order structure function of the streamwise velocity component, applying the Taylor hypothesis to convert the measured time series into spatial velocity variations:

\begin{equation}
D_{11}(\tau)=\left\langle\left[u^{\prime}(t+\tau)-u^{\prime}(t)\right]^{2}\right\rangle
\end{equation}

\begin{equation}
D_{11}(r)=C_{2} \epsilon^{2 / 3} r^{2 / 3}
\label{D11}
\end{equation}

With the Kolmogorov prediction for the second-order structure function in the inertial range (equation \ref{D11}), where $C_2$ is a constant of around 2 in high-Reynolds number turbulence \citep{Saddoughi94}, we can estimate the turbulence dissipation rate $\epsilon=\left(D_{11} /\left(C_{2} r^{2 / 3}\right)\right)^{3 / 2}$. Furthermore, we calculated the Kolmogorov time and length scale, $\tau_{\eta}=(v / \epsilon)^{1 / 2}$ and $\eta=\left(v^{3} / \epsilon\right)^{1 / 4}$, the Taylor microscale, $\lambda=u_\textnormal{rms}(15 v / \epsilon)^{1 / 2}$, and the Reynolds number, $\Rey_{\lambda}=u_{\mathrm{rms}} \lambda / \nu$, where $\nu$ is the kinematic viscosity of air, and $u_{\mathrm{rms}}$ is the root mean square (r.m.s.) of the velocity fluctuations $u^{\prime}$.

\section {Results}\label{sec:results}
\subsection {Snow particle acceleration and Stokes number}\label{sec:31}
\begin{figure}
  \centerline{\includegraphics[scale = 0.85]{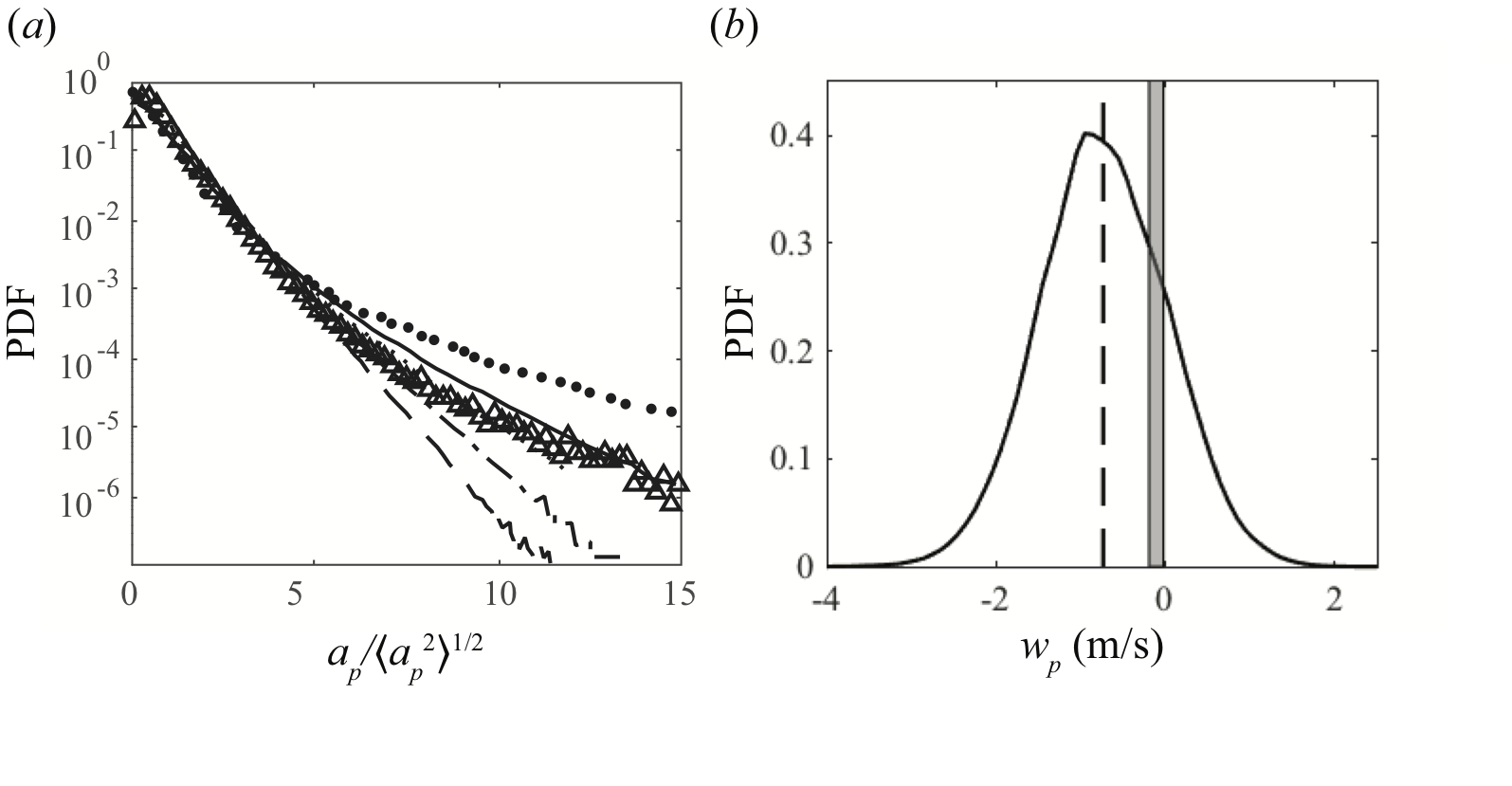}}
  \caption{(\textit{a}) PDFs of in-plane snow particle acceleration from PTV (triangles), compared to $\Stk$ = 0 from \citet{Mordant04} (dots), and \citet{Bec06} ($\Stk$ = 0.16, solid line; $\Stk$ = 0.37, dotted line; $\Stk$ = 1.01, dash-dotted line; $\Stk$ = 2.03, dashed line); (\textit{b}) comparison of the measured distribution (solid line) and average (dashed line) of settling velocity with the estimated range of still-air terminal velocities (grey region).}
\label{fig:3}
\end{figure}

The snow particle acceleration and vertical velocity obtained from PTV analysis are evaluated in this section. Figure \ref{fig:3}a shows the probability density function (PDF) of the fluctuations of the snow particle acceleration normalized by their r.m.s. value. The PDF is compared with data from previous laboratory experiments and numerical simulations of tracers and inertial particles in isotropic turbulence \citep{Mordant04,Bec06,Ayya06}. In figure \ref{fig:3}a, the exponential tail of the in-plane acceleration PDF curve of snow particles lies in between the curves with Stokes numbers of 0.16 and 1.01 from \citet{Bec06}, while a comparison of streamwise acceleration with \citet{Ayya06}, in a similar boundary layer flow, seem to narrow the range to 0.09-0.15. As discussed in \citet{Nemes17}, the acceleration kurtosis manifests the tendency of inertial particles to experience only a portion of the high acceleration events sustained by fluid parcels. The direct numerical simulation (DNS) by \citet{Ireland16} showed that the kurtosis of acceleration becomes insensitive to the change of Reynolds number with $\Stk > 0.1$ and $\Rey_{\lambda} > 398$ (e.g., as $\Rey_{\lambda}$ changes from 398 to 597 at $\Stk = 0.1$, the kurtosis of acceleration increases only 3$\%$, and the change becomes smaller at higher $\Stk$). Therefore, following the reasoning in our previous studies \citep{Nemes17,Li21}, we extend the comparison of the acceleration PDFs to the atmospheric turbulence case with high $\Rey_{\lambda}$ investigated here, and conservatively estimate the Stokes number in the range of 0.09-1.01.

With the estimated $\Stk$, the aerodynamic particle response time of the observed snow particles is predicted to be in the range from 1.7 ms to 19.6 ms, where $\tau_p = \Stk\cdot\tau_{\eta}$, leading to a still-air terminal velocity defined by ${\it W_p}= \tau_p\cdot{g}$ and estimated between 0.02 m/s and 0.19 m/s. The estimated Stokes number indicates a near critical condition ($\Stk\sim\textit{O}(1)$), anticipating the occurrence of preferential concentration (clustering) and sweeping, as well as enhanced settling velocity. In figure 3b, we compare the vertical velocity distribution (solid line) from PTV (the average vertical velocity ($\langle w_p\rangle$) of 0.73 m/s is indicated as a dashed line) with the estimated terminal velocity range accounting for the uncertainty in Stokes number (grey region). The increase is evident and multifold (around seven times larger on average, $\langle w_p\rangle / \overline{W}_{p}$). This enhancement is consistent with what has been observed in the previous study by \citet{Nemes17} (around three times enhancement on average). However, since our estimated range for $\Stk$ is closer to the critical value, the observed enhancement here is higher. 

Note that the corresponding particle Reynolds number $\Rey = {\langle w_p\rangle}D/\nu$ based on the measured settling velocity and particle size is $\sim$16.8, implying that a non-Stokesian drag correction is required. Due to the disk-like shape of the observed snow particles, the Schiller-Neumann approach based on the Reynolds-corrected sphere drag is not recommended, which in part justifies the estimation of Stokes number from the acceleration PDF with no explicit dependency on density, and size. The only option to account for snow morphology is to use the semi-empirical $\chi$ number approach proposed by \citet{Bohm89}, corrected by \citet{Heymsfield10}, and also employed in \citet{Nemes17}. The resulting parameterization leads to $\chi = 997$ and a drag coefficient of $C_\textnormal{D}=3.53$, which is not unusual given the relatively low particle Reynolds number \citep{Westbrook17}. It is important to stress that the $\chi$ number accounts for the snow morphology effects on drag \citep{Garrett15, Dunnavan19}, not necessarily for the effect of ambient turbulence, which is the main point of this work.

\subsection {Preferential distribution of snow particle concentration}\label{sec:32}
\begin{figure}
  \centerline{\includegraphics[scale = 0.85]{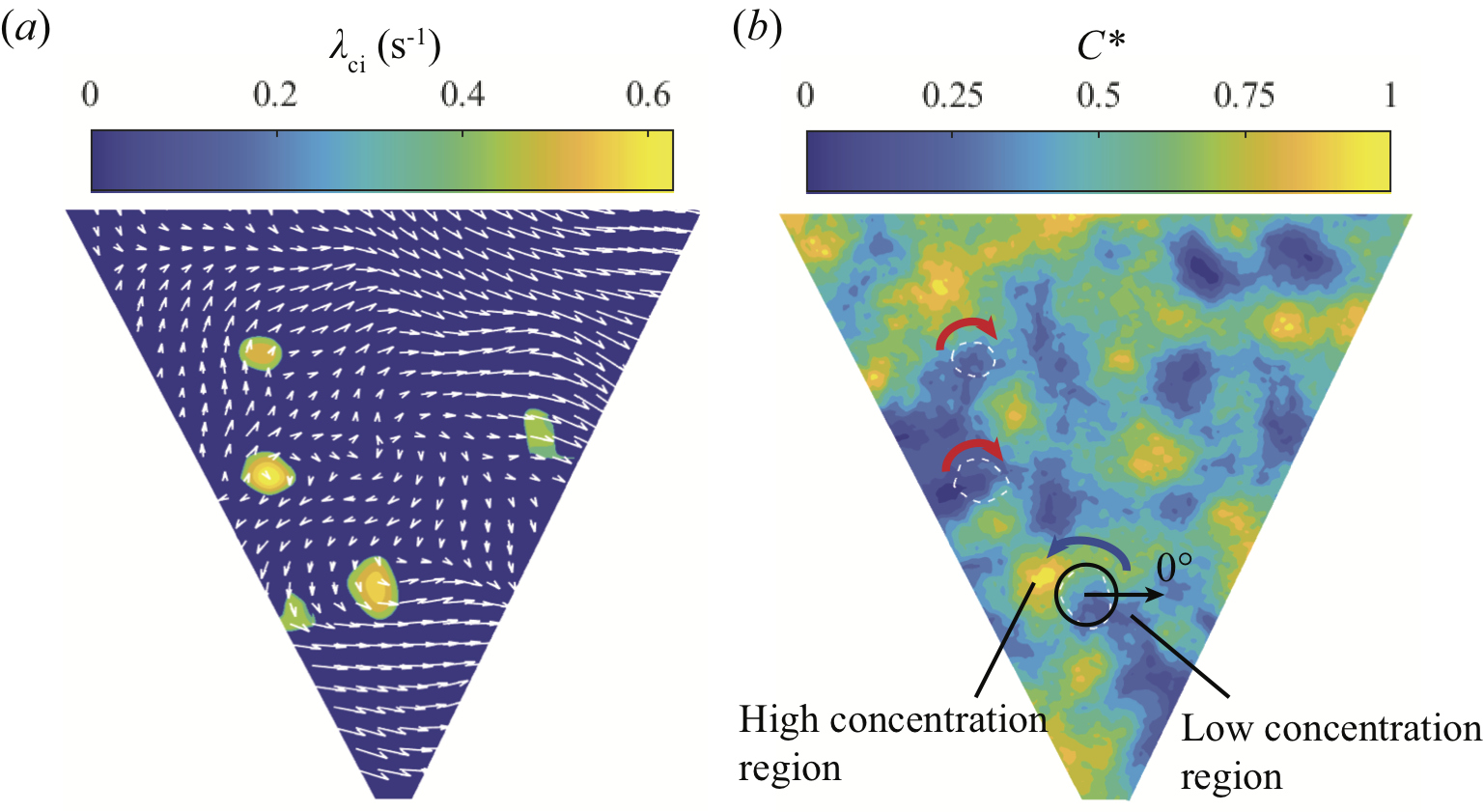}}
  \caption{(\textit{a}) Swirling strength contour of an instantaneous sample (threshold 0.4 $\textnormal{s}^{-1}$ applied) with relative flow velocity vectors (white arrows, subtracted by the convective velocity of 3.4 m/s of the prograde vortex at the center left of the field of view). (\textit{b}) Corresponding snow particle concentration colormap ($C^{*}$ is the relative concentration) with vortices detected from (\textit{a}). Note that the white dashed lines represent vortex boundaries, red arrows indicate prograde vortices and blue arrow indicates the retrograde vortex, black circle and arrow define the coordinate system of the local vortex.}
\label{fig:4}
\end{figure}

Snow particle concentration around vortices in the flow is first evaluated using the SLPIV data. As shown in an instantaneous flow sample (figure \ref{fig:4}a), the vortices are identified using the swirling strength derived from the velocity fields of SLPIV as described in detail in \cref{sec:21}. Subsequently, the particle number concentration within and around the vortices is estimated using the image intensity ($I(x,z,t)$) of the SLPIV data. The estimation of concentration using image intensity is supported by \citet{Raffel18} which shows that the image intensity of PIV is proportional to the concentration of particles with the same averaged diameters. However, factors such as stochastic light attenuation by particles within the light sheet and in between the light sheet and the cameras, as well as the power fluctuation of the search light might lead to non-linear relationship between the light intensity and local particle concentration \citep{Kalt07, Banko19}. Nevertheless, due to relatively low volume fraction, the light attenuation by particles between the light sheet and the cameras is not inferred to be dominant as compared to the other two factors. Furthermore, to minimize the spatial and temporal non-uniformity in background image intensity due to the decay of light intensity with height and its fluctuation over time, relative concentration $C^{*}=I(x, z, t) / I(x, z)_{\mathrm{10s,avg}}$ is defined according to \citet{Li21}, where $I(x, z)_{\mathrm{10s,avg}}$ is an average of the intensity of images recorded in a 10 s moving window. Figure \ref{fig:4}b shows the relative concentration of the flow sample corresponding to figure \ref{fig:4}a. In particular, we observe that snow concentration is low within the vortex cores. The phenomenon is considered to be a result of the inertia bias \citep{Maxey87}, i.e., inertial particle trajectories are biased towards the region of low vorticity. Remarkably, for all three strong vortices (i.e., vortices with high swirling strength values) highlighted in the figure, including both prograde and retrograde vortices, the particle concentration is preferentially higher on the downward flow side of the vortices (referred to as preferential concentration hereafter).

It is worth noting that the relative concentration map in figure \ref{fig:4}b seems to suggest the clustering of snow particles in the atmospheric surface layer similar to those studied in our previous work \citep{Li21}. However, the clusters in figure 4b are on smaller scales and do not exhibit a clear preferential orientation in comparison to those presented in \citet{Li21} with the January 2019 dataset. The difference is mainly caused by relatively lower turbulence and snow concentration in the current cases which lead to a weaker interaction between particles and turbulence. In addition, since the current study focuses on elucidating the connection between the vortical structures in the turbulent flow and settling of individual snow particle, the PTV was designed to have a more focused field of view (3 m $\times$ 5 m, smaller than the integral scale), limiting our ability to quantify the large-scale clusters that extend beyond our region of interest.

\begin{figure}
  \centerline{\includegraphics[scale = 0.85]{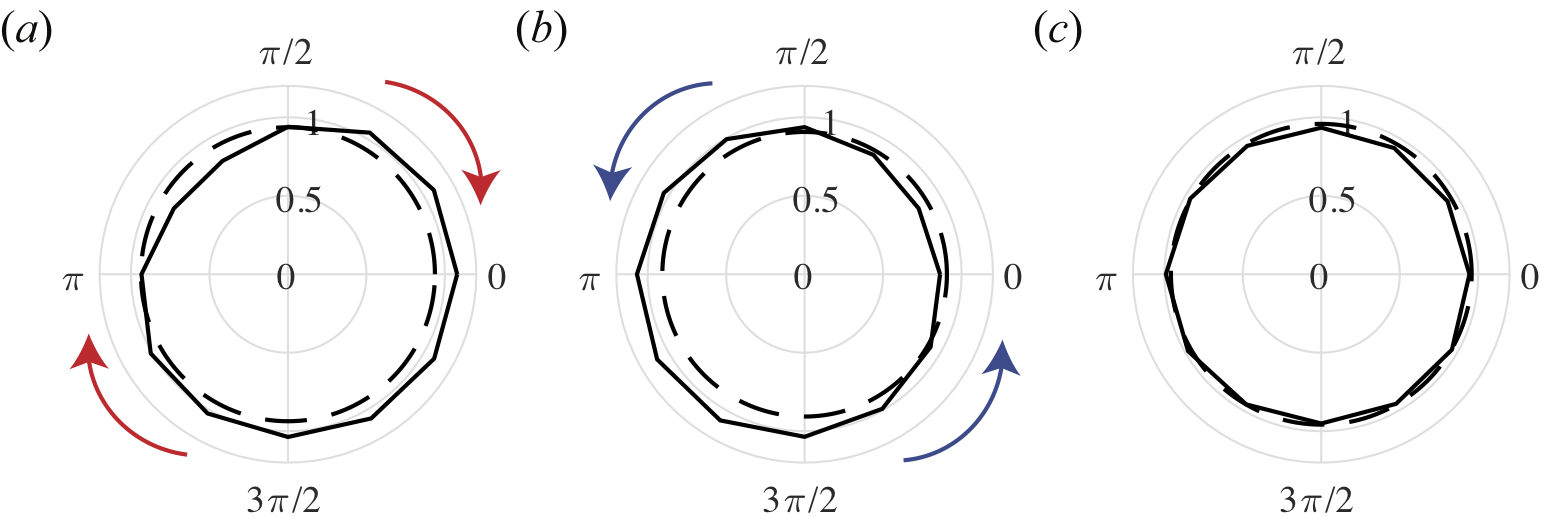}}
  \caption{Comparison between the ensemble-averaged concentration from the SLPIV dataset in the central region (dashed lines), defined as the region within half $R_{\mathrm{eff}}$ from the vortex center where $R_{\mathrm{eff}}$ is the effective radius of the vortices, and in the proximity (solid lines), defined as the region from $0.7R_{\mathrm{eff}}$ to $1.4R_{\mathrm{eff}}$ from the center spanning a circumferential angle range of $\pi / 6$, for (\textit{a}) prograde vortices, (\textit{b}) retrograde vortices, and (\textit{c}) reference background. For each prograde/retrograde vortex, the reference background is defined as a circular region that has a radius equal to the $R_{\mathrm{eff}}$ of the prograde/retrograde vortex with a center at a randomly selected location in the SLPIV measurement field of view.}
\label{fig:5}
\end{figure}

To substantiate the observation of preferential concentration associated with presence of vortices in the flow, the ensemble-averaged concentration contours are calculated for prograde (28,700 prograde vortices identified) and retrograde vortices (9,700 retrograde vortices), respectively. As shown in figure \ref{fig:5}, the averaged concentration is determined for both the central region defined as the region within half effective radius ($R_{\mathrm{eff}}=\sqrt{A / \pi}$, where \textbf{\textit{A}} is the area of the vortex region detected through the swirling strength criterion) from the center of the vortex, and in the rest of proximity defined as the region from $0.7R_{\textnormal{eff}}$ to $1.4R_{\textnormal{eff}}$ from the center. The latter is divided into twelve angular sectors with angle of $\pi / 6$. For both prograde (figure \ref{fig:5}a) and retrograde vortices (figure \ref{fig:5}b), it is observed that the particle concentration is higher on the downward side, than that on the upward side and in the center of vortices. For reference, the averaged concentration distribution in the background (figure \ref{fig:5}c) does not exhibit any appreciable preferential concentration. For each prograde/retrograde vortex, the reference background is defined as a circular region that has a radius equal to the $R_{\textnormal{eff}}$ of the prograde/retrograde vortex with a center at a randomly selected location in the SLPIV measurement field of view. In addition, the averaged particle concentration is relatively higher at the bottom of the downward side for retrograde vortices, possibly due to gravity and downward fluid motion causing a stronger preferential concentration effect. But for prograde vortices, relative concentration is more uniformly distributed on the downward side. Such a discrepancy can be explained by the different organizations of prograde and retrograde vortices in the atmospheric surface layer. Specifically, the prograde vortices tend to form in packets \citep{Chris01}, predominantly located in the internal shear layers in atmospheric surface layer, the interaction between snow particles around a certain prograde vortex with adjacent vortices may smooth out the particle concentration on the downward side.

It is well known that preferential concentration in turbulent flows is a multi-scale phenomenon. As shown in \citep{Baker17}, particles start clustering at the Kolmogorov scale when they preferentially sample the high strain regions in the flow. As the clusters yield larger response time than that of individual particles, they can subsequently interact with larger scale flow structures and grow in size up to the integral scale (see examples in \citet{Li21}). However, due to the limit of the SLPIV resolution, we can only resolve vortices above a certain size ($\sim$ 66 cm) in our measurements. Therefore, the preferential concentration statistics shown in figure \ref{fig:5} are captured only by sampling large, energetic vortices leaving a signature several times larger than the Taylor microscale in our coarsely resolved turbulent flow. Nevertheless, the estimated Stokes number in \cref{sec:31} (with the upper range close to the critical condition) suggests strong interaction between the particles and flow structure at the Kolmogorov scales, causing preferential concentration and clustering that cannot be resolved with the current SLPIV measurement. We thus acknowledge resolving a limited range of scale contributing to particle clustering, but capturing the key mechanism governing preferential sweeping at the resolved scale.

\begin{figure}
  \centerline{\includegraphics[scale = 0.85]{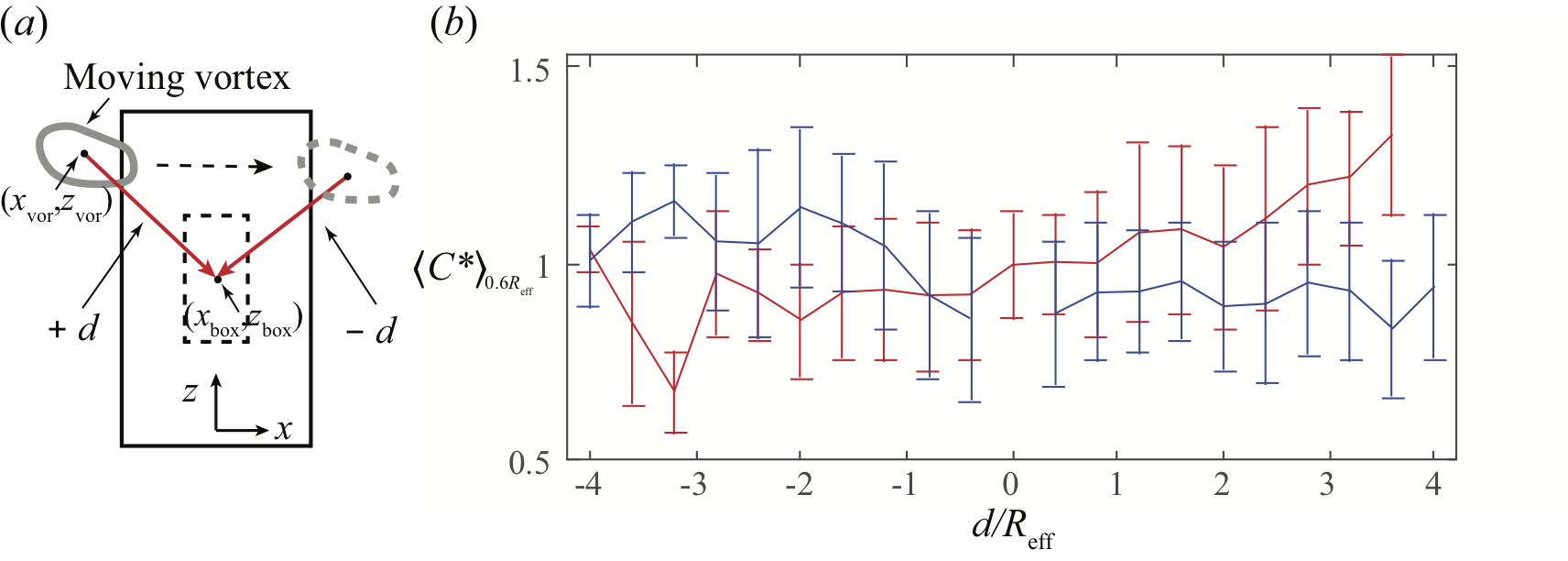}}
  \caption{(\textit{a}) A schematic showing the box counting process to determine the change of snow particle concentration due to the presence of a vortex (gray solid and dashed contours) in the PTV sample area (black solid line). The vortex is determined from the corresponding SLPIV measurement at the same time instant as the PTV. A fixed counting box (black dashed line, 100 $\times$ 200 $\textnormal{pixel}^2$) is selected at the center of the PTV area. The relative displacement (red arrow) between the vortex center ($x_{\mathrm{vor}}$, $z_{\mathrm{vor}}$) and box center ($x_{\mathrm{box}}$, $z_{\mathrm{box}}$) is defined by equation \ref{dboxvor}. (\textit{b}) The ensemble-averaged particle concentration ($\left\langle C^{*}\right\rangle_{0.6 R_{\mathrm{eff}}}$) at different relative displacements with respect to the vortex center. In total, 550 prograde (red line) and 380 retrograde (blue line) vortices are selected for the ensemble average, respectively. The bin of relative displacement ($\Delta d$)  used in ensemble average has a width of $0.6R_{\mathrm{eff}}$ and spaced $0.4R_{\mathrm{eff}}$ from adjacent bins ($\sim$33$\%$ overlap). The bin size and spacing are determined to ensure sufficient statistical convergence of the data.}
\label{fig:6}
\end{figure}

Furthermore, we examine the images from the PTV measurements in which individual snow particles can be counted and tracked within and around the vortices (see movies 1 and 2 in the supplementary material), supporting the observation of preferential concentration based on the change in the intensity of images from SLPIV. We first select the vortices determined from the corresponding SLPIV measurements at the same time instant as the PTV with overlapping field of view. Most of these vortices have an equivalent diameter $\gtrsim$1.3 m and are only partially inside the PTV images as they move across the small (relative to SLPIV) sample area of PTV. As a result, the previous method (that for SLPIV) for estimating particle concentration cannot be applied. Instead, as illustrated in figure \ref{fig:6}a, a box counting method is used to determine the change of snow particle concentration due to the presence of vortices in the PTV sample area (note that the uncertainty of particle concentration associated with the changing light sheet thickness within the PTV field is estimated to be less than 9$\%$). A fixed counting box (100 $\times$ 200 $\textnormal{pixel}^2$) is selected at the center of the PTV area. When the vortex appears in the sample area of PTV, the total number of snow particles in the box is counted. The relative displacement (\textit{d}) between the vortex center ($x_{\mathrm{vor}}$, $z_{\mathrm{vor}}$) and box center ($x_{\mathrm{box}}$, $z_{\mathrm{box}}$) is calculated as:
\begin{equation}
d=\frac{x_{\mathrm{box}}-x_{\mathrm{vor}}}{\left|x_{\mathrm{box}}-x_{\mathrm{vor}}\right|} \sqrt{\left(x_{\mathrm{box}}-x_{\mathrm{vor}}\right)^{2}+\left(z_{\mathrm{box}}-z_{\mathrm{vor}}\right)^{2}}
\label{dboxvor}
\end{equation}
Particularly, the sign of the relative displacement indicates which side the particles are located with respect to the moving vortex. Similar to that in the SLPIV data processing, to account for the variation of snow particle concentration in the background, the relative concentration for PTV data ($C^{*}$) is estimated using total particle number counts in the box at each time instant divided by the averaged total particle counts for the time duration during which each vortex is present in the sample area of PTV. Subsequently, the ensemble averaged $C^{*}$ in a bin of a width of $0.6R_{\textnormal{eff}}$ ($\left\langle C^{*}\right\rangle_{0.6 R_{\mathrm{eff}}}$) at different relative displacements is determined for prograde (550 in total) and retrograde (380 in total) vortices, respectively (figure \ref{fig:6}b). As the figure shows, for prograde vortices, the $\left\langle C^{*}\right\rangle_{0.6 R_{\mathrm{eff}}}$ is evidently higher on the downward side of the vortices (positive $d$). For retrograde vortices, the $\left\langle C^{*}\right\rangle_{0.6 R_{\mathrm{eff}}}$ also yields larger values on the downward side (negative $d$), though the difference in $\left\langle C^{*}\right\rangle_{0.6 R_{\mathrm{eff}}}$ between two sides appears to be smaller than that observed for prograde vortices. To determine the significance of the difference in $\left\langle C^{*}\right\rangle_{0.6 R_{\mathrm{eff}}}$, we conduct a t-test to the concentration distribution at the two sides of vortices: the particle concentration on the downward side is in general 13$\%$-22$\%$ higher (at 95$\%$ confidence level) than that at the upward side for both prograde and retrograde vortices. These trends (see also the supplementary material) provide further support of the preferential concentration. Nevertheless, due to the limited number of vortices that can be simultaneously detected by SLPIV and PTV, the standard deviation of the $\left\langle C^{*}\right\rangle_{0.6 R_{\mathrm{eff}}}$ presented in the current analysis is considerable. In addition, we would like to point out that the statistical signature of preferential concentration observed in PTV is likely to be underestimated potentially due to the relatively large counting box size used in the analysis in comparison to the size of vortices.

\subsection {Enhanced settling velocity due to preferential sweeping}
\begin{figure}
  \centerline{\includegraphics[scale = 0.85]{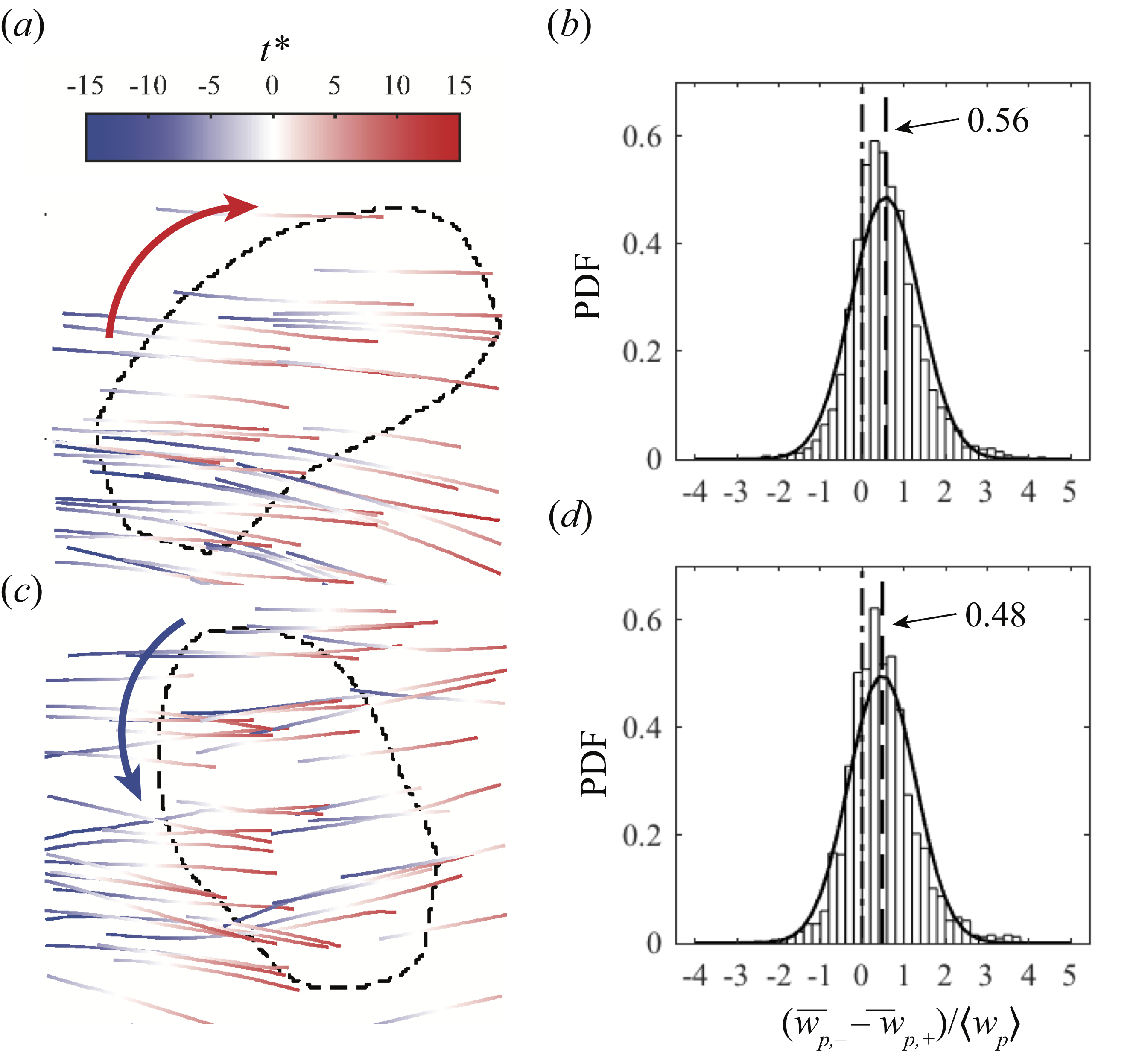}}
\caption{(\textit{a}, \textit{c}) Samples of snow particle trajectories around a prograde vortex (\textit{a}) and a retrograde vortex (\textit{c}) (see also movie 1 and 2 uploaded in the supplementary material). Black dashed lines represent vortex boundaries, and trajectories are colored based on the dimensionless times $t^{*}$, defined as the difference between the timestamps of snow particles and that of a selected vortex at one time instant normalized by the Kolmogorov time scale. (\textit{b}, \textit{d}) Histograms of the differences between the settling velocities on the downward sides and upward sides for (\textit{b}) prograde and (\textit{d}) retrograde vortices (A total of 4300 prograde and 1700 retrograde vortices are identified), normalized by the average vertical velocity $\langle w_{p}\rangle$ of snow particles tracked with zeros marked by the dash-dotted lines, and the dashed lines represent the average velocity difference in each case. The bin size is chosen to be one fifth of $\langle w_{p}\rangle$. The histograms are compared with the Gaussian distributions marked by the solid lines.}
\label{fig:7}
\end{figure}

In this subsection, we investigate how the settling velocity of snow particles is influenced by the presence of the vortical structures in the flow. Consistent with the method presented in the last section, the vortices are first identified using SLPIV data. Subsequently, the particle trajectories around these identified vortices are extracted from PTV for the following analysis. For the prograde vortex (figure \ref{fig:7}a), the settling velocity of particles increases when moving toward the downward side (right side in the sample) of the vortex. Similarly, for the retrograde vortex (figure \ref{fig:7}c), the settling of snow particles slows down; some particles are even lifted upward, as they travel to the upward side (right side in the sample) of the vortex. Both cases illustrate clearly higher settling velocities of the snow particles situating on the downward side of vortices.

To further substantiate these observations, the average vertical accelerations conditioned on the downward ($\overline{a_{p,-}}$) and the upward ($\overline{a_{p,+}}$) sides of prograde and retrograde vortices are calculated, and the histograms of settling velocity difference between the downward ($\overline{w_{p,-}}$) and upward ($\overline{w_{p,+}}$) sides of vortices, i.e., $\overline{w_{p,-}}-\overline{w_{p,+}}$, normalized by the ensemble average snow particle vertical velocity ($\langle w_{p}\rangle=0.73 \mathrm{~m} / \mathrm{s}$) are presented for all prograde (figure \ref{fig:7}b) and retrograde (figure \ref{fig:7}d) vortices, respectively. The $\overline{a_{p,-}}$, $\overline{a_{p,+}}$, $\overline{w_{p,-}}$ and $\overline{w_{p,+}}$ are calculated by averaging the vertical velocities of particles within boxes ranging from $0.5R_{\mathrm{eff}}$ to $1.5R_{\mathrm{eff}}$ to the center of vortices in the \textit{x} direction and covering the whole diameter ($2R_{\mathrm{eff}}$) in the vertical direction at the two sides of vortices. Specifically, for prograde vortices, the two conditionally averaged accelerations are $\overline{a_{p,-}} = -0.16 \pm 2.20 \mathrm{m/s^{2}}$ on the downward side and $\overline{a_{p,+}} = 0.0065 \pm 2.54 \mathrm{m/s^{2}}$ on the upward side. While for retrograde vortices, $\overline{a_{p,-}} = -0.33 \pm 2.93 \mathrm{m/s^{2}}$ and $\overline{a_{p,+}} = 0.12 \pm 2.09 \mathrm{m/s^{2}}$. These conditional averages support the fact that particles on the downward side of vortices would accelerate with the flow and particles falling on the upward side would decelerate, or even be lifted up. Note that the variability in acceleration (e.g. the standard deviation) is much higher than that for the settling velocity due to the higher order derivative in the acceleration calculation. Moreover, the settling velocity differences display a near Gaussian distribution with the mean value above zero. As compared to Gaussian distribution with the same mean value and standard deviation, the PDFs of settling velocity difference exhibit higher probability near the mean values and heavier tails on the right side for both prograde and retrograde vortices, indicating statistically higher settling velocities on the downward side of vortices. Specifically, about 78$\%$ of the prograde vortices yield a higher settling velocity on the downward side with an average settling velocity differences of 0.56$\langle w_p\rangle$, and the proportion is about 73$\%$ for the retrograde vortices with an average settling velocity difference of 0.48$\langle w_p\rangle$. Note that the total number of vortices identified in figure \ref{fig:7} is larger than that for particle concentration in figure \ref{fig:6}. It is because we identify individual vortex from the PTV field for settling velocity analysis, while vortices selected for preferential concentration in figure \ref{fig:6} are tracked across the region of interest.

However, not every single occasion is observed with higher settling velocity at the downward side of vortices, due to the fact that the vortices detected in the current study are planar projections of highly complex three-dimensional vortices. In addition, the snow particles interacting with one vortex can also be affected by adjacent vortices in atmosphere, which are usually weaker and less appreciable in the SLPIV data as compared to the vortices analyzed. Nevertheless, we observe an increasing percentage of events showing such a trend when we sample vortices with higher swirling strength and the percentage for prograde vortices is consistently higher than that for retrograde vortices. Such discrepancies are likely due to the difference in self-organization characteristics for prograde and retrograde vortices (i.e., prograde vortices are found to be predominantly located in the proximity of internal shear layers as shown in the field PIV measurement in the atmospheric surface layer by \citet{Heisel18}). Therefore, prograde vortices could have a cumulative and stronger effect on the enhanced settling of nearby snow particles compare to retrograde vortices. Nevertheless, with the observed preferential concentration from \cref{sec:32} and statistically higher settling velocity on the downward side of the vortices, we conclude that under near critical conditions ($\Stk\sim\textit{O}(1)$) observed in our field measurements, preferential sweeping plays a significant role in controlling the settling velocity of hydrometeors in the atmospheric turbulence.

\section{Conclusions and discussion}\label{sec:con}
In this paper, we present the first field study of snow settling dynamics based on simultaneous measurements of the atmospheric flow field using a super-large-scale particle image velocimetry (SLPIV) and snow particle trajectories using particle tracking velocimetry (PTV) within the SLPIV sample area. Our results reveal the direct linkage among the coherent vortex structures in the atmospheric turbulence, and the concentration distribution and settling dynamics of snow particles in the field. Specifically, we observe a settling velocity enhancement of around seven folds on average compared to the estimated still-air terminal velocity. This value is larger than what has been observed in our previous field study \citep{Nemes17}, potentially due to the fact that the Stokes number associated with the snow particles in the present deployment is closer to the critical condition ($\Stk\sim\textit{O}(1)$). Using the SLPIV, we are able to detect the strong vortices present in the atmospheric turbulent flow, and we show that the snow concentration (represented by the variation in the particle image intensity in SLPIV) is preferentially higher on the downward side for both prograde and retrograde vortices in the flow. This observation is further substantiated by counting individual snow particles around the vortices using PTV data. The result indicates an average of 18$\%$ higher concentration on the downward side of the detected strong vortices present in our study. In addition, the samples of snow particle trajectories around vortices from PTV demonstrate that snow particles accelerate as they move toward the downward side of vortices and decelerate or even are lifted upward when traveling to the upward side. Based on the histograms of snow settling velocity from PTV, we show that the snow particles on the downward side of vortices yield a statistically higher settling velocity than that at the upward side, with an average difference of about 52$\%$ of the mean settling velocity.

Our results provide direct evidence and underlying mechanisms for the preferential concentration and preferential sweeping during snow particles settling in the atmospheric surface layer. While the presented results focus on the quantification of particle-turbulence interaction mechanisms at the scale of individual vortices, we recognize that atmospheric turbulence in the near-neutrally stratified boundary layer in our study contains vortices over a broad range of scales and intensity. Therefore, we hypothesize that the preferential paths of highly concentrated particles along layers of vortices are likely to produce a cumulative effect on enhancing the settling of the snow particles. This conceptual framework has not been considered in current snow settling models. Under this framework, our observed results of preferential concentration and preferential sweeping may be critical to inform a stochastic model to reproduce the observed fall speed under specific micrometeorological conditions. By incorporating the framework, it may be possible to model the settling of snow particles starting from the still-air terminal velocity with specific drag coefficient and sampling a population of vortices consistent with the inertial range in the air column. Under the assumption of accelerated preferential paths, each snow particle has a large probability to sample a number of vortices, and the effects of both the drag force and vortex flow sweeping will progressively enhance the settling velocity to a certain value. With this perspective, it is reasonable to hypothesize that the multifold increase in observed settling velocity compared with the still-air terminal velocity, as well as its large variability, is a result of the cumulative effect of particles settling through the turbulent air column occupied by many vortices. In addition, the disparity in settling velocities observed on the downward and upward sides of vortices of ~0.4 m/s might also be directly affected by the azimuthal velocity of the flow around vortices, manifested in vertical velocity fluctuations that are more readily available from the measured flow field.

In the end, we acknowledge the uncertainties involved in our concentration measurement using box counting from the PTV data. Such uncertainties are largely caused by the limited data of synchronized SLPIV and PTV measurements from our field deployment. Nevertheless, the main observations related to the snow particle concentration and settling dynamics present in our study are still statistically significant. A second relevant uncertainty is in the estimate of the snow particle velocity in still air. While the Stokes number range is conservatively defined, it would still be important to provide a direct estimate of the snow particle density, combining measurements of single particle volume with the weighing of particle ensemble in time. We expect that more converged trends can be obtained with an increasing number of deployments. In particular, we expect to extend our current field PTV to a three-dimensional imaging system using multiple cameras.  Such upgraded system will allow us to quantify, more accurately, the particle concentration and distribution, particle setting kinematics (e.g., curvature of the trajectories) and dynamics (e.g., acceleration, inertial response) associated with the presence of three-dimensional vortex structures in the atmospheric turbulence.

\bibliographystyle{jfm}

\begin{thebibliography}{99}

\expandafter\ifx\csname natexlab\endcsname\relax
\def\natexlab#1{#1}\fi
\expandafter\ifx\csname selectlanguage\endcsname\relax
\def\selectlanguage#1{\relax}\fi

\bibitem[Abraham and Hong(2020)]{Abraham20}
{\sc Abraham, A. and Hong, J. } 2020 {Dynamic wake modulation induced by utility-scale wind turbine operation}, {\it Appl. Energy}, {\bf 257}, 114003.

\bibitem[Adrian \etal(2000)]{Adrian00}
{\sc Adrian, R. J., Christensen, K. T. and Liu, Z. C. } 2000 {Analysis and interpretation of instantaneous turbulent velocity fields}, {\it Exp. Fluids}, {\bf 29} (3), 275-290.

\bibitem [Aliseda \etal(2002)]{Aliseda02}
{\sc Aliseda, A., Cartellier, A., Hainaux, F. and Lasheras, J. C. } 2002 {Effect of preferential concentration on the settling velocity of heavy particles in homogeneous isotropic turbulence}, {\it J. Fluid Mech.}, {\bf 468}, 77-105.

 \bibitem[Ayyalasomayajula \etal(2006)] {Ayya06}
{\sc Ayyalasomayajula, S., Gylfason, A., Collins, L. R., Bodenschatz, E. and Warhaft, Z.} 2006 {Lagrangian measurements of inertial particle accelerations in grid generated wind tunnel turbulence}, {\it Phys. Rev. Lett.}, {\bf 97} (14), 144507.

\bibitem [Baker \etal(2017)]{Baker17}
{\sc Baker, L., Frankel, A., Mani, A. and Coletti, F. } 2017 {Coherent clusters of inertial particles in homogeneous turbulence}, {\it J. Fluid Mech.}, {\bf 833}, 364-398.

\bibitem[Balachandar and Eaton(2010)]{Balachandar10}
 {\sc Balachandar, S. and Eaton, J. K. } 2010  {Turbulent dispersed multiphase flow}, {\it Annu. Rev. Fluid Mech.}, {\bf 42}, 111-133.
 
 \bibitem[Banko \etal(2019)]{Banko19}
 {\sc Banko, A. J., Villafañe, L., Kim, J. H., Esmaily, M. and Eaton, J. K.} 2019  {Stochastic modeling of direct radiation transmission in particle-laden turbulent flow}, {\it J. Quant. Spectrosc. Radiat. Transf.}, {\bf 226}, 1-18.

\bibitem[Bec \etal(2006)] {Bec06}
  {\sc Bec, J., Biferale, L., Boffetta, G., Celani, A., Cencini, M., Lanotte, A., Musacchio, S. and Toschi, F.} 2006 {Acceleration statistics of heavy particles in turbulence}, {\it J. Fluid Mech.}, {\bf 550}, 349-358.

\bibitem [Böhm(1989)] {Bohm89}
 {\sc  Böhm, H. P.} 1989 {A general equation for the terminal fall speed of solid hydrometeors},  {\it J. Atmos. Sci.}, {\bf 46} (15), 2419-2427.
 
 
 \bibitem [Christensen and Adrian(2001)]{Chris01}
{\sc Christensen, K. T. and Adrian, R. J. } 2001 {Statistical evidence of hairpin vortex packets in wall turbulence}, {\it J. Fluid Mech.}, {\bf 431}, 433-443.
 
 \bibitem [Crocker and Grier(1996)] {Crocker96}
 {\sc  Crocker, J. C. and Grier, D. G.} 1996 {Methods of digital video microscopy for colloidal studies},  {\it J. Colloid Interface Sci.}, {\bf 179} (1), 298-310.

\bibitem [Dasari \etal(2019)] {Dasari19}
{\sc Dasari, T., Wu, Y., Liu, Y. and Hong, J.} 2019 {Near-wake behaviour of a utility-scale wind turbine},  {\it J. Fluid Mech.}, {\bf 859}, 204-246.

\bibitem [Dunnavan \etal(2019)] {Dunnavan19}
 {\sc  Dunnavan, E. L., Jiang, Z., Harrington, J. Y., Verlinde, J., Fitch, K. and Garrett, T. J.} 2019 {The shape and density evolution of snow aggregates},  {\it J. Atmos. Sci.}, {\bf 76} (12), 3919-3940.

\bibitem [Durán \etal(2011)] {Duran11}
{\sc Durán, O., Claudin, P. and Andreotti, B.} 2011 {On aeolian transport: Grain-scale interactions, dynamical mechanisms and scaling laws},  {\it Aeolian Res.}, {\bf 3} (3), 243-270.

\bibitem [Elghobashi and Truesdell(1992)] {Elghobashi92}
{\sc Elghobashi, S. and Truesdell, G. C.} 1992 {Direct simulation of particle dispersion in a decaying isotropic turbulence},  {\it J. Fluid Mech.}, {\bf 242}, 655-700.

\bibitem [Falkinhoff \etal(2020)] {Falkinhoff20}
{\sc Falkinhoff, F., Obligado, M., Bourgoin, M. and Mininni, P. D.} 2020 {Preferential concentration of free-falling heavy particles in turbulence},  {\it Phys. Rev. Lett.}, {\bf 125} (6), 064504.

\bibitem [Ferrante and Elghobashi(2003)] {Ferrante03}
{\sc Ferrante, A. and Elghobashi, S.} 2003 {On the physical mechanisms of two-way coupling in particle-laden isotropic turbulence},  {\it Phys. Fluids}, {\bf 15} (2), 315-329.

\bibitem [Garrett \etal(2015)] {Garrett15}
{\sc Garrett, T. J., Yuter, S. E., Fallgatter, C., Shkurko, K., Rhodes, S. R. and Endries, J. L.} 2015 {Orientations and aspect ratios of falling snow},  {\it Geophys. Res. Lett.}, {\bf 42} (11), 4617-4622.

\bibitem [Good \etal(2014)] {Good14}
{\sc Good, G. H., Ireland, P. J., Bewley, G. P., Bodenschatz, E., Collins, L. R. and Warhaft, Z.} 2014 {Settling regimes of inertial particles in isotropic turbulence},  {\it J. Fluid Mech.}, {\bf 759} (R3).

\bibitem [Heisel \etal(2018)] {Heisel18}
{\sc Heisel, M., Dasari, T., Liu, Y., Hong, J., Coletti, F. and Guala, M.} 2018 {The spatial structure of the logarithmic region in very-high-Reynolds-number rough wall turbulent boundary layers},  {\it J. Fluid Mech.}, {\bf 857}, 704-747.

\bibitem [Heymsfield and Westbrook(2010)] {Heymsfield10}
{\sc Heymsfield, A. J. and Westbrook, C. D.} 2010 {Advances in the estimation of ice particle fall speeds using laboratory and field measurements},  {\it J. Atmos. Sci.}, {\bf 67} (8), 2469-2482.

\bibitem [Högström \etal(2002)] {Hogstrom02}
{\sc Högström, U., Hunt, J. C. R. and Smedman, A. S.} 2002 {Theory and measurements for turbulence spectra and variances in the atmospheric neutral surface layer},  {\it Bound.-Layer Meteorol.}, {\bf 103} (1), 101-124.

\bibitem [Hong \etal(2014)] {Hong14}
{\sc Hong, J., Toloui, M., Chamorro, L.P., Guala, M., Howard, K., Riley, S., Tucker, J. and Sotiropoulos, F.} 2014 {Natural snowfall reveals large-scale flow structures in the wake of a 2.5-MW wind turbine},  {\it Nat. Commun.}, {\bf 5}, 4216.

\bibitem [Ireland \etal(2016)] {Ireland16}
{\sc Ireland, P. J., Bragg, A. D. and Collins, L. R.} 2016 {The effect of Reynolds number on inertial particle dynamics in isotropic turbulence. Part 1. Simulations without gravitational effects},  {\it J. Fluid Mech.}, {\bf 796}, 617-658.

\bibitem [Kalt \etal(2007)] {Kalt07}
{\sc Kalt, P. A., Birzer, C. H. and Nathan, G. J.} 2007 {Corrections to facilitate planar imaging of particle concentration in particle-laden flows using Mie scattering, Part 1: Collimated laser sheets.},  {\it Appl. Opt.}, {\bf 46} (23), 5823-5834.

\bibitem [Li \etal(2021)] {Li21}
{\sc Li, C., Lim, K., Berk, T., Abraham, A., Heisel, M., Guala, M., Coletti, F. and Hong, J.} 2021 {Settling and clustering of snow particles in atmospheric turbulence},  {\it J. Fluid Mech.}, {\bf 912} (A49).

\bibitem [Mallery \etal(2020)] {Mallery20}
{\sc Mallery, K., Shao, S. and Hong, J.} 2020 {Dense particle tracking using a learned predictive model},  {\it Exp. Fluids}, {\bf 61} (10), 1-14.

\bibitem [Maxey(1987)] {Maxey87}
{\sc Maxey, M. R.} 1987 {The gravitational settling of aerosol particles in homogeneous turbulence and random flow fields.},  {\it J. Fluid Mech.}, {\bf 174}, 441-465.

\bibitem [Maxey and Riley(1983)] {Maxey83}
{\sc Maxey, M. R. and Riley, J. J.} 2020 {Equation of motion for a small rigid sphere in a nonuniform flow},  {\it Phys. Fluids}, {\bf 26} (4), 883-889.

\bibitem [Mordant \etal(2004)] {Mordant04}
{\sc Mordant, N., Crawford, A. M. and Bodenschatz, E.} 2004 {Experimental Lagrangian acceleration probability density function measurement},  {\it Physica} D, {\bf 193} (1-4), 245-251.

\bibitem [Nemes \etal(2017)] {Nemes17}
{\sc Nemes, A., Dasari, T., Hong, J., Guala, M. and Coletti, F.} 2017 {Snowflakes in the atmospheric surface layer: observation of particle–turbulence dynamics},  {\it J. Fluid Mech.}, {\bf 814}, 592-613.

\bibitem [Nielsen(1993)] {Nielson93}
{\sc Nielsen, P.} 1993 {Turbulence effects on the settling of suspended particles},  {\it J. Sediment. Res.}, {\bf 63} (5), 835-838.

\bibitem[Ouellette \etal(2006)]{Ouellette06}
{\sc Ouellette, N. T., Xu, H. and Bodenschatz, E. } 2006 {A quantitative study of three-dimensional Lagrangian particle tracking algorithms}, {\it Exp. Fluids}, {\bf 40} (2), 301-313.

\bibitem [Petersen \etal(2019)] {Petersen19}
{\sc Petersen, A. J., Baker, L. and Coletti, F.} 2019 {Experimental study of inertial particles clustering and settling in homogeneous turbulence},  {\it J. Fluid Mech.}, {\bf 864}, 925-970.

\bibitem [Raffel \etal(2018)] {Raffel18}
{\sc Raffel, M., Willert, C. E., Scarano, F., Kähler, C. J., Wereley, S. T. and Kompenhans, J.} 2018 {\it Particle Image Velocimetry: A Practical Guide}. Springer.

\bibitem [Rosa \etal(2016)] {Rosa16}
{\sc Rosa, B., Parishani, H., Ayala, O. and Wang, L. P.} 2016 {Settling velocity of small inertial particles in homogeneous isotropic turbulence from high-resolution DNS},  {\it Int. J. Multiph. Flow}, {\bf 83}, 217-231.

\bibitem [Saddoughi and Veeravalli(1994)] {Saddoughi94}
{\sc Saddoughi, S. G. and Veeravalli, S. V.} 1994 {Local isotropy in turbulent boundary layers at high Reynolds number},  {\it J. Fluid Mech.}, {\bf 268}, 333-372.

\bibitem [Shaw(2003)] {Shaw03}
{\sc Shaw, R. A.} 2003 {Particle-turbulence interactions in atmospheric clouds},  {\it Annu. Rev. Fluid Mech.}, {\bf 35} (1), 183-227.

\bibitem [Stull(1988)] {Stull88}
{\sc Stull, R. B.} 1988 {\it An introduction to boundary layer meteorology}. Kluwer Academic Publishers.

\bibitem [Toloui \etal(2014)] {Toloui14}
{\sc Toloui, M., Riley, S., Hong, J., Howard, K., Chamorro, L. P., Guala, M. and Tucker, J.} 2014 {Measurement of atmospheric boundary layer based on super-large-scale particle image velocimetry using natural snowfall},  {\it Exp. Fluids}, {\bf 55} (5), 1-14.

\bibitem [Tom and Bragg(2019)] {Tom19}
{\sc Tom, J. and Bragg, A.} 2019 {Multiscale preferential sweeping of particles settling in turbulence},  {\it J. Fluid Mech.}, {\bf 872}, 995-995.

\bibitem [Tooby \etal(1977)] {Tooby77}
{\sc Tooby, P. F., Wick, G. L. and Isaacs, J. D.} 1977 {The motion of a small sphere in a rotating velocity field: a possible mechanism for suspending particles in turbulence},  {\it J. Geophys. Res.}, {\bf 82} (15), 2096-2100.

\bibitem [Tropea \etal(2007)] {Tropea07}
{\sc Tropea, C., Yarin, A. L. and Foss, J. F.} 2007 {Springer handbook of experimental fluid mechanics}. Berlin: Springer.

\bibitem [Vaillancourt and Yau(2000)] {Vaillancourt00}
{\sc Vaillancourt, P. A. and Yau, M. K.} 2000 {Review of particle-turbulence interactions and consequences for cloud physics},  {\it Bull. Am. Meteorol. Soc.}, {\bf 81} (2), 285-298.

\bibitem [Wang and Maxey(1993)] {Wang93}
{\sc Wang, L. P. and Maxey, M. R.} 1993 {Settling velocity and concentration distribution of heavy particles in homogeneous isotropic turbulence},  {\it J. Fluid Mech.}, {\bf 256}, 27-68.

\bibitem [Westbrook and Sephton(2017)] {Westbrook17}
{\sc Westbrook, C. D. and Sephton, E. K.} 2017 {Using 3‐D‐printed analogues to investigate the fall speeds and orientations of complex ice particles},  {\it Geophys. Res. Lett.}, {\bf 44} (15), 7994-8001.

\bibitem [Yang and Lei(1998)] {Yang98}
{\sc Yang, C. Y. and Lei, U.} 1998 {The role of the turbulent scales in the settling velocity of heavy particles in homogeneous isotropic turbulence},  {\it J. Fluid Mech.}, {\bf 371}, 179-205.

\bibitem [Zeugin \etal(2020)] {Zeugin20}
{\sc Zeugin, T., Krol, Q., Fouxon, I. and Holzner, M.} 2020 {Sedimentation of snow particles in still air in stokes regime},  {\it Geophys. Res. Lett.}, {\bf 47} (15), e2020GL087832.

\bibitem [Zhou \etal(1999)] {Zhou99}
{\sc Zhou, J., Adrian, R. J., Balachandar, S. and Kendall, T. M.} 1999 {Mechanisms for generating coherent packets of hairpin vortices in channel flow},  {\it J. Fluid Mech.}, {\bf 387}, 353-396.

\end{thebibliography}

\end{document}